\documentclass{article}
\usepackage{spconf,amsmath,graphicx}
\usepackage{makecell}
\usepackage{graphicx}
\usepackage{subcaption}
\usepackage{float}
\usepackage{booktabs}
\usepackage{xcolor}
\usepackage{amssymb}

\title{An Attention-Enhanced Network with Joint Dehazing and Retinex-Based Enhancement for Underwater Images}
\name{Sahana Ray, Bibhabasu Debnath, and Sanjay Ghosh, Senior Member, IEEE}
\address{Department of Electrical Engineering, IIT Kharagpur, WB 721302, India}
\begin{document}

\maketitle

\begin{abstract}
Underwater images suffer from severe wavelength-dependent light absorption and scattering, and turbidity due to suspended particles, degrading visual quality for applications in autonomous underwater vehicles (AUVs), marine biology, archaeology, and offshore infrastructure inspection. Classical IFM inadequately capture nonlinear underwater light behavior, while purely data-driven methods lack physical interpretability. This paper proposes a three-stage network named \textbf{ADR}, that extends the underwater image formation model with additional terms to perform underwater dehazing, followed by Retinex-based enhancement and attention-enabled U-Net++ refinement. Experiments on UIEB and UFO-120 benchmark datasets demonstrate competitive performance with state-of-the-art methods. 
\end{abstract}


\section{Introduction}
\label{sec:intro}

Light propagation underwater differs from atmospheric conditions due to water's optical properties~\cite{duntley1963light,schechner2006recovery}. Long wavelengths attenuate rapidly, causing color distortion and low contrast, while scattering introduces haze. Still, underwater images remain crucial for ecological and biological research, including species monitoring and environmental assessment.

Dehazing and color correction~\cite{zhang2022underwater,zhang2023underwater} have emerged as central problems underwater image enhancement (UIE). 
We consider an image formation model (IFM) as follows:
\begin{equation}
I(x) = J(x) \cdot t(x) + A \cdot (1 - t(x)),
\label{eq:classical_model}
\end{equation}
where $x \in \Omega$ denotes the spatial pixel location, $I(x) \in \mathbb{R}^3$ is the observed degraded image, $J(x) \in \mathbb{R}^3$ is the scene radiance, $t(x) \in \mathbb{R}$ is the medium transmission map, $A \in \mathbb{R}^3$ denotes the global background light, and $\cdot$ denotes element-wise multiplication~\cite{jaffe2002computer,narasimhan2000chromatic,fattal2008single}. He et al.~\cite{he2010single} introduced the dark channel prior (DCP), observing that in most haze-free images, local patches contain at least one color channel with very low intensity. However, DCP fails underwater since the red channel remains severely attenuated everywhere. Drews et al.~\cite{drews2013transmission} thus restricted transmission estimation to blue and green channels. Akkaynak and Treibitz~\cite{akkaynak2018revised} proposed a revised IFM with explicit wavelength dependence, and Peng~\emph{et al.}~\cite{peng2017underwater} incorporated blurriness-based depth estimation.

Retinex theory~\cite{land1971lightness} models an image as the pixel-wise product of illumination and reflectance:
\begin{equation}
I(x) = L(x) \cdot R(x),
\label{eq:retinex}
\end{equation}
where $L(x) \in \mathbb{R}$ represents the spatially varying illumination and $R(x) \in \mathbb{R}^3$ denotes the reflectance capturing intrinsic object properties~\cite{ghosh2019fast,ghosh2019fast1}. Zhang et al.~\cite{zhang2017underwater} used multi-scale Retinex in LAB space, and Galdran et al.~\cite{galdran2018duality} established a relation between dehazing and Retinex decomposition. 


Deep learning approaches typically achieve superior results. Water-Net~\cite{li2019underwater} employs an encoder-decoder with skip connections; SCNet~\cite{fu2022underwater} uses spatial and channel normalization within a U-Net backbone; Espinosa et al.~\cite{espinosa2023efficient} developed a U-Net with DWT skip connections and CBAM; and Chen et al.~\cite{chen2021underwater} proposed lightweight CNNs for joint transmission and background light estimation.

In this paper, we present a three-stage architecture for underwater image enhancement. The first stage employs a multi-branch network to estimate depth, transmission map, background light, and additional parameters enabling dehazing through an extended IFM incorporating turbidity and noise terms. The second stage adopts Retinex-based decomposition for adaptive gamma correction of illumination and reflectance refinement. The third stage employs a U-Net++-based enhancement network with self-attention to integrate multi-scale contextual information from all preceding stages. 
Based on three core themes in our work -- \textit{attention}, \textit{dehazing}, and \textit{retinex}, we refer our proposed method as \textbf{ADR}. 

The remainder of this paper is organized as follows. Section~\ref{sec:method} details the proposed ADR framework, including the network architecture and loss function. Section~\ref{sec:exp} presents results and comparisons, while Section~\ref{sec:conclusion} concludes the paper.

\section{Proposed Method}
\label{sec:method}

\begin{figure*}[!h]
\centering
\includegraphics[width=0.9\textwidth]{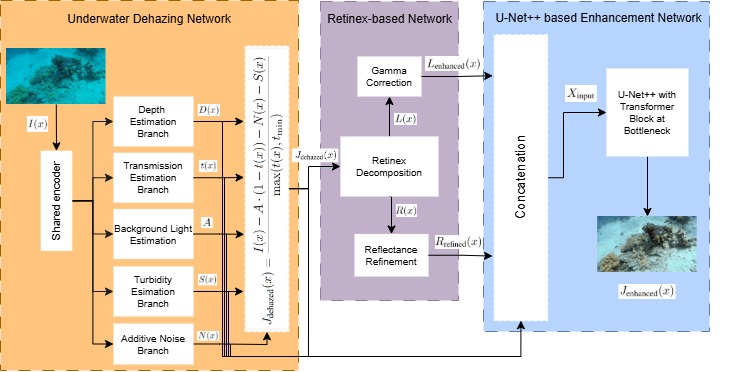}
\caption{Three-stage underwater image enhancement architecture. Stage 1: A shared encoder feeds five parallel branches to recover the dehazed image. Stage 2: Retinex decomposition with learned gamma correction on illumination map and refinement on reflectance map. Stage 3: A U-Net++ enhancement network with transformer bottleneck, which produces final output image.}
\label{fig:arch}
\end{figure*}

The proposed ADR framework employs a three-stage architecture that addresses the specific degradation challenges encountered in underwater imaging, as illustrated in Figure~\ref{fig:arch}.

\subsection{Enhanced Physics-Guided Dehazing}

The first stage implements an extended physics-based correction model. To account for additional degradation factors in underwater environments, the model in~\eqref{eq:classical_model} is extended as:
\begin{equation}
I(x) = J(x) \cdot t(x) + A \cdot (1 - t(x)) + N(x) + S(x),
\label{eq:extended_model}
\end{equation}
where $N(x) \in \mathbb{R}^3$ denotes additive noise and $S(x) \in \mathbb{R}^3$ represents turbidity-induced scattering components.
A shared encoder with three convolutional layers (32, 64, 128 filters) and max pooling feeds five parallel branches. The depth branch produces a single-channel map $D(x) \in \mathbb{R}^{1 \times H \times W}$, the transmission branch outputs $t(x) \in \mathbb{R}^{3 \times H \times W}$ representing wavelength-dependent attenuation, the background light branch predicts $A \in \mathbb{R}^{3}$ via pooling and FC layers.

The noise branch with Conv($128 \rightarrow 256 \rightarrow 128 \rightarrow 3$), ReLU activations, and bilinear upsampling predicts:
\begin{equation}
N(x) = \alpha_N \hat{N}(x) \cdot \exp(-\gamma_D \cdot D(x)),
\label{eq:additive}
\end{equation}
where $\hat{N}(x) \in \mathbb{R}^{3 \times H \times W}$ is the predicted noise map, $\alpha_N \in \mathbb{R}$ is a learnable noise scaling factor, and $\gamma_D \in \mathbb{R}$ is a learnable depth attenuation factor. The formulation reflects the observation that noise increases with depth. The turbidity branch with the same Conv structure predicts:
\begin{equation}
S(x) = \beta_T \tanh(\hat{S}(x)) \cdot (1 - t(x)) \cdot D(x),
\label{eq:turbidity}
\end{equation}
where $\hat{S}(x) \in \mathbb{R}^{3 \times H \times W}$ is the predicted turbidity map and $\beta_T \in \mathbb{R}$ is a learnable scaling factor. The factor $(1-t(x))$ emphasizes high-attenuation regions and $D(x)$ accounts for depth-dependent scattering accumulation, while $\tanh(\cdot)$ constrains the response to a stable range. The dehazed image is recovered by inverting \eqref{eq:extended_model}:
\begin{equation}
J_{\text{dehazed}}(x) = \frac{I(x) - A \cdot (1 - t(x)) - N(x) - S(x)}{\max(t(x),\, t_{\min})},
\label{eq:enhanced_dehazing}
\end{equation}
where $t_{\min} = 0.1$ prevents numerical instability.

\subsection{Retinex-Based Decomposition}

The second stage decomposes the dehazed image into illumination and reflectance components following \eqref{eq:retinex} using a lightweight encoder-decoder: a shared encoder ($3 \rightarrow 32 \rightarrow 64$ channels) feeds separate decoders for the illumination map $L(x) \in [0,1]^{1 \times H \times W}$ and reflectance map $R(x) \in [0,1]^{3 \times H \times W}$, both with sigmoid activation. The illumination undergoes learned gamma correction:
\begin{equation}
L_{\text{enhanced}}(x) = L(x)^{\gamma(x)},
\label{eq:gamma_correction}
\end{equation}
where $\gamma(x) \in [0.5, 1.5]$ is predicted by a convolutional layer with sigmoid activation plus 0.5 offset, brightening underexposed regions while preserving well-lit areas. The reflectance passes through a refinement layer ($3 \rightarrow 3$) with sigmoid activation, producing $R_{\text{refined}}(x) \in [0,1]^{3 \times H \times W}$. Both $L_{\text{enhanced}}(x)$ and $R_{\text{refined}}(x)$ are concatenated with other intermediate features and passed to the third stage.

\subsection{Main Enhancement Network}

The third stage employs a U-Net++ topology with transformer-based attention. The network receives a 20-channel concatenated input from all previous stages:
\begin{equation}
\begin{aligned}
X_{\text{input}} = [&I(x),\; J_{\text{dehazed}}(x),\; L_{\text{enhanced}}(x),\; R_{\text{refined}}(x),\\
                   &D(x),\; t(x),\; N(x),\; S(x)].
\end{aligned}
\label{eq:concatenation}
\end{equation}
The encoder progressively increases channel capacity to 64, 128, 256, and 512 across four levels following the nested dense skip connection pattern of U-Net++~\cite{zhou2018nested}, with a four-head transformer self-attention block at the bottleneck. The decoder reconstructs features through dense multi-scale skip connections, and a final $1\times1$ convolutional layer projects to the output RGB image $J_{\text{enhanced}}(x)$ with sigmoid activation.

\subsection{Composite Loss Function}
To supervise the image enhancement network, a composite loss function is employed that integrates multiple complementary objectives. The overall training objective is:
\begin{equation}
\begin{aligned}
\mathcal{L}_{\text{total}} =
&\lambda_1 \mathcal{L}_{L1} +
\lambda_2 \mathcal{L}_{\text{SSIM}} +
\lambda_3 \mathcal{L}_{\text{perc}} +
\lambda_4 \mathcal{L}_{\text{dehaze}} +
\lambda_5 \mathcal{L}_{\text{retinex}} ,
\end{aligned}
\label{eq:total_loss}
\end{equation}
where $\lambda_i$ denotes the weighting coefficient for each loss term.

\noindent1. \textit{L1 Loss:} The L1 loss provides pixel-wise supervision between the enhanced image and the ground truth:
\begin{equation}
\mathcal{L}_{L1} =
\frac{1}{N} \sum_{i=1}^{N}
\left| \hat{J}(i) - J_{\text{GT}}(i) \right| ,
\label{eq:l1_loss}
\end{equation}
where $N$ denotes the total number of pixels, $\hat{J}$ is the final enhanced output, and $J_{\text{GT}}$ is the ground truth.

\noindent2. \textit{SSIM Loss:} To maintain structural integrity, an SSIM-based loss explicitly models perceptual similarity:
\begin{equation}
\mathcal{L}_{\text{SSIM}} = 1 - \text{SSIM}(\hat{J},\, J_{\text{GT}}) ,
\label{eq:ssim_loss}
\end{equation}
where SSIM is computed as:
\begin{equation}
\text{SSIM}(x, y) =
\frac{(2\mu_x\mu_y + C_1)(2\sigma_{xy} + C_2)}
{(\mu_x^2 + \mu_y^2 + C_1)(\sigma_x^2 + \sigma_y^2 + C_2)} ,
\label{eq:ssim_formula}
\end{equation}
with $\mu_x$, $\mu_y$ the local means, $\sigma_x^2$, $\sigma_y^2$ the local variances, $\sigma_{xy}$ the local covariance, and $C_1$ and $C_2$ are stability constants.

\noindent3. \textit{Perceptual Loss:} This loss measures feature-level discrepancies using an unweighted VGG-style network $\phi(\cdot)$:
\begin{equation}
\mathcal{L}_{\text{perc}} =
\left\| \phi(\hat{J}) - \phi(J_{\text{GT}}) \right\|_2^2 .
\label{eq:perc_loss}
\end{equation}

\noindent4. \textit{Dehaze Supervision Loss:} To provide intermediate supervision on the dehazing stage, the recovered scene radiance $J$ is directly compared against the ground truth:
\begin{equation}
\mathcal{L}_{\text{dehaze}} =
\frac{1}{N} \sum_{i=1}^{N}
\left| J(i) - J_{\text{GT}}(i) \right| ,
\label{eq:dehaze_loss}
\end{equation}
where $J$ is the output of the physics-guided stage (Stage~1) before 
Retinex decomposition and UNet++ refinement.

\noindent5. \textit{Retinex Reconstruction Loss:} To enforce consistency of the Retinex decomposition, the product of the estimated illumination map $L$ and reflectance map $R$ is penalised against the ground truth:
\begin{equation}
\mathcal{L}_{\text{retinex}} =
\frac{1}{N} \sum_{i=1}^{N}
\left| L(i) \cdot R(i) - J_{\text{GT}}(i) \right| ,
\label{eq:retinex_loss}
\end{equation}
where $L$ is broadcast to match the spatial dimensions of $R$ prior to multiplication.

\section{Experiments}
\label{sec:exp}

\subsection{Dataset and Implementation Details}

The proposed model was trained and evaluated on the UIEB (Underwater Image Enhancement Benchmark) dataset~\cite{li2019underwater}, a widely adopted benchmark containing 890 paired underwater images, partitioned into 700 training and 190 test images. Additionally, the model was trained and evaluated on the UFO-120~\cite{islam2020semantic} dataset containing 1,500 training and 120 test paired samples. All images were resized to $256 \times 256$. The three-stage network was jointly trained over 100 epochs on an NVIDIA GeForce RTX 3060 GPU with 12GB memory. Adam optimizer was used with initial learning rate $10^{-4}$ and batch size 8. The coefficients in~\eqref{eq:total_loss} empirically set to $\lambda_1{=}1.0$, $\lambda_2{=}0.5$, $\lambda_3{=}0.1$, $\lambda_4{=}0.3$, and $\lambda_5{=}0.2$.

\subsection{Quantitative Results}

Model performance was quantitatively evaluated using three widely adopted metrics: peak signal-to-noise ratio (PSNR) for pixel-level reconstruction accuracy, structural similarity index measure (SSIM) for perceptual similarity \eqref{eq:ssim_formula}, and learned perceptual image patch similarity (LPIPS)~\cite{zhang2018unreasonable} for measuring perceptual distance using deep features.

Quantitative comparisons on the UIEB and UFO-120 datasets are reported in Tables~\ref{tab:uiebd} and~\ref{tab:ufo120}, respectively. On UIEB, the proposed method achieves an SSIM of 0.9040, PSNR of 22.97~dB, and LPIPS of 0.1175, outperforming all compared methods in SSIM and LPIPS. Semi-UIR obtains a marginally higher PSNR by exploiting additional unpaired data via contrastive semi-supervised learning, whereas the proposed method trains exclusively on paired images. The large margins over classical methods such as IBLA~\cite{peng2017underwater} and PCDE~\cite{zhang2023underwater} confirm that physics-based priors alone are insufficient for real-world underwater degradations, while consistent LPIPS gains reflect perceptually sharper outputs which can be partially attributed to perceptual loss training. On the additional UFO-120 benchmark (Table~\ref{tab:ufo120}), the proposed method achieves 29.18~dB PSNR, 0.9249 SSIM, and 0.1012 LPIPS, outperforming all baselines across all metrics. During inference, the proposed method runs at 0.53~s per image on UIEB (Table~\ref{tab:inference}), higher than lightweight baselines such as Chen~\emph{et al.}~\cite{chen2021underwater} (0.09~s) due to the complex architecture.

\begin{table}[t]
\centering
\caption{Performance comparisons on the UIEB dataset. Best results are marked in \textbf{bold}.}
\vspace{-0.3cm}
\label{tab:uiebd}
\setlength{\tabcolsep}{7pt}
\begin{tabular}{lccc}
\toprule
Method & SSIM $\uparrow$ & PSNR $\uparrow$ & LPIPS $\downarrow$ \\
\midrule
PCDE~\cite{zhang2023underwater} & 0.7032 & 15.83 & 0.3498\\
IBLA~\cite{peng2017underwater} & 0.5733 & 14.39 & 0.4299 \\
Water-Net~\cite{li2019underwater} & 0.8303 & 19.31 & 0.2016 \\
Chen~\emph{et al.}~\cite{chen2021underwater} & 0.8770 & 19.37 & 0.1947 \\
Ucolor~\cite{li2021underwater} & 0.8412 & 21.97 & 0.1945 \\
Shallow-UWnet~\cite{naik2021shallow} & 0.8496 & 19.48 & 0.1828 \\
SCNet~\cite{fu2022underwater} & 0.8625 & 22.08 & 0.1936 \\
Espinosa~\emph{et al.}~\cite{espinosa2023efficient} & 0.8802 & 20.92 & -- \\
NU2Net~\cite{guo2023underwater} & 0.8606 & 19.80 & 0.1833 \\
Semi-UIR~\cite{huang2023contrastive} & 0.8680 & \textbf{23.63} & 0.1200 \\
\midrule
\textbf{ADR} & \textbf{0.9040} & 22.97 & \textbf{0.1175} \\
\bottomrule
\end{tabular}
\end{table}

\begin{table}[!h]
\centering
\caption{Performance comparisons on the UFO-120 dataset.}
\vspace{-0.3cm}
\label{tab:ufo120}
\setlength{\tabcolsep}{7pt}
\begin{tabular}{lccc}
\toprule
Method & SSIM $\uparrow$ & PSNR $\uparrow$ & LPIPS $\downarrow$ \\
\midrule
Water-Net~\cite{li2019underwater} & 0.7330 & 23.12 & 0.3112 \\
Shallow-UWnet~\cite{naik2021shallow} & 0.8292 & 25.04 & 0.2480 \\
SCNet~\cite{fu2022underwater} & 0.8689 & 27.12 & 0.2539 \\
NU2Net~\cite{guo2023underwater} & 0.8540 & 25.70 & 0.1612 \\
Semi-UIR~\cite{huang2023contrastive} & 0.8837 & 28.41 & 0.2160 \\
\midrule
\textbf{ADR} & \textbf{0.9249} & \textbf{29.18} & \textbf{0.1012} \\
\bottomrule
\end{tabular}
\end{table}
\begin{table}[!h]
\centering
\caption{Inference time (s) comparison on UIEB dataset.}
\vspace{-0.3cm}
\label{tab:inference}
\setlength{\tabcolsep}{4pt}
\begin{tabular}{cccccccc}
\toprule
& \cite{zhang2023underwater} & \cite{peng2017underwater} & \cite{li2019underwater} & \cite{chen2021underwater} & \cite{fu2022underwater} & \cite{espinosa2023efficient} & ADR\\
\midrule
Time(s) & 0.32 & 38.71 & 0.61 & \textbf{0.09} & 0.45 & 0.33 & 0.53 \\
\bottomrule
\end{tabular}
\end{table}

\subsection{Ablation Study}
\begin{table}[!h]
\centering
\vspace{-0.3cm}
\caption{Ablation study on UIEB test set (256$\times$256).}
\vspace{-0.3cm}
\label{tab:ablation}
\setlength{\tabcolsep}{7pt}
\begin{tabular}{lccc}
\toprule
Model & PSNR $\uparrow$ & SSIM $\uparrow$ & LPIPS $\downarrow$ \\
\midrule
\textbf{Full model}        & 22.97 & \textbf{0.9040} & 0.1175 \\
w/o Turbidity term      & 22.90 & 0.8997 & 0.1183 \\
w/o Noise term        & \textbf{23.11} & 0.9026 & 0.1168 \\
w/o Retinex stage        & 22.97 & 0.9031 & 0.1159 \\
w/o U-Net++ stage         & 22.85 & 0.9033 & \textbf{0.1146} \\
w/o $\mathcal{L}_{\text{perc}}$      & 22.87 &  0.9026  & 0.1191  \\
w/o $\mathcal{L}_{\text{SSIM}}$         & 22.57 & 0.8982 & 0.1223\\
\bottomrule
\end{tabular}
\vspace{-0.2cm}
\end{table}
Systematic ablations are performed on UIEB test set for images of dimension 256$\times$256 to evaluate the contributions of critical components of the model. Removing the turbidity branch makes all metrics drop, although not significantly. Without the noise branch, the model performs slightly better in terms of PSNR and LPIPS, while SSIM reduces; hence, depth-dependent noise serves as a structural guidance rather than the primary reconstruction driver in the scattering-dominated UIEB dataset. Although the impact of Retinex branch is not dominant in the ablation, it contributes to the improvement of perceptual quality, confirmed by an improvement in SSIM. Without the U-Net++ stage, there is consistent degradation across PSNR and SSIM, confirming its essential role in the model. Further, $\mathcal{L}_{\text{SSIM}}$ contributes significantly more to the network compared to $\mathcal{L}_{\text{SSIM}}$.

\subsection{Qualitative Analysis}

\begin{figure*}[t]
    \centering
    \makebox[0.18\textwidth][c]{Input}
    \hspace{0.01\textwidth}
    \makebox[0.18\textwidth][c]{Chen et al.}
    \hspace{0.01\textwidth}
    \makebox[0.18\textwidth][c]{SCNet}
    \hspace{0.01\textwidth}
    \makebox[0.18\textwidth][c]{ADR}
    \hspace{0.01\textwidth}
    \makebox[0.18\textwidth][c]{Ground Truth}

    \vspace{0.4em}

    \includegraphics[width=0.18\textwidth]{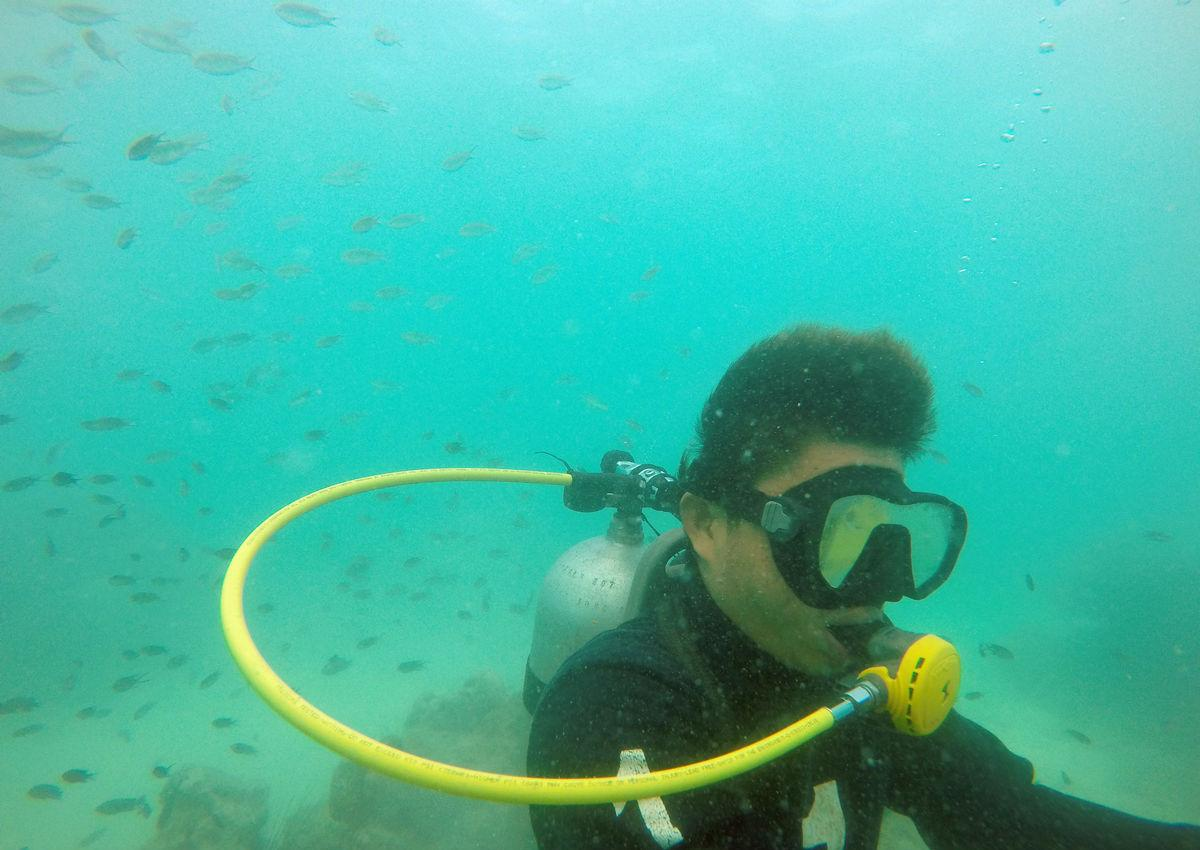}\hspace{0.01\textwidth}
    \includegraphics[width=0.18\textwidth]{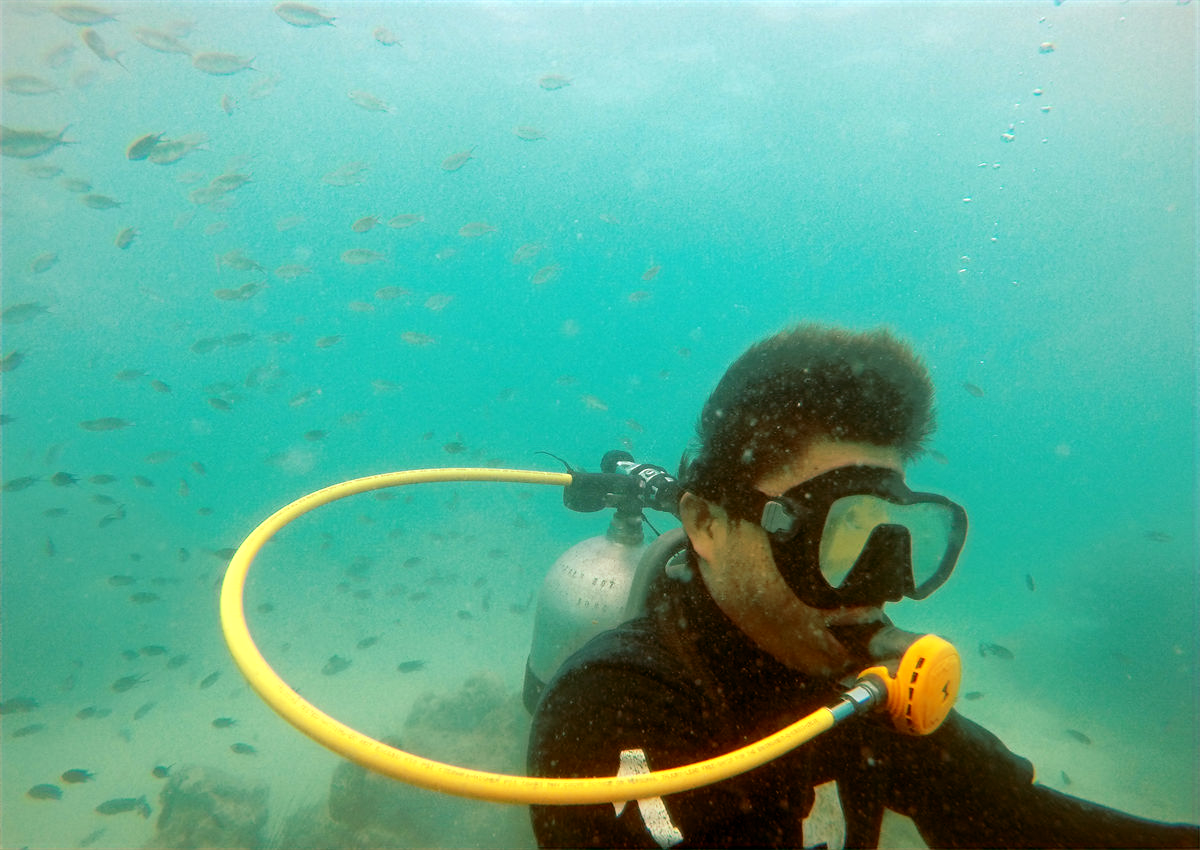}\hspace{0.01\textwidth}
    \includegraphics[width=0.18\textwidth]{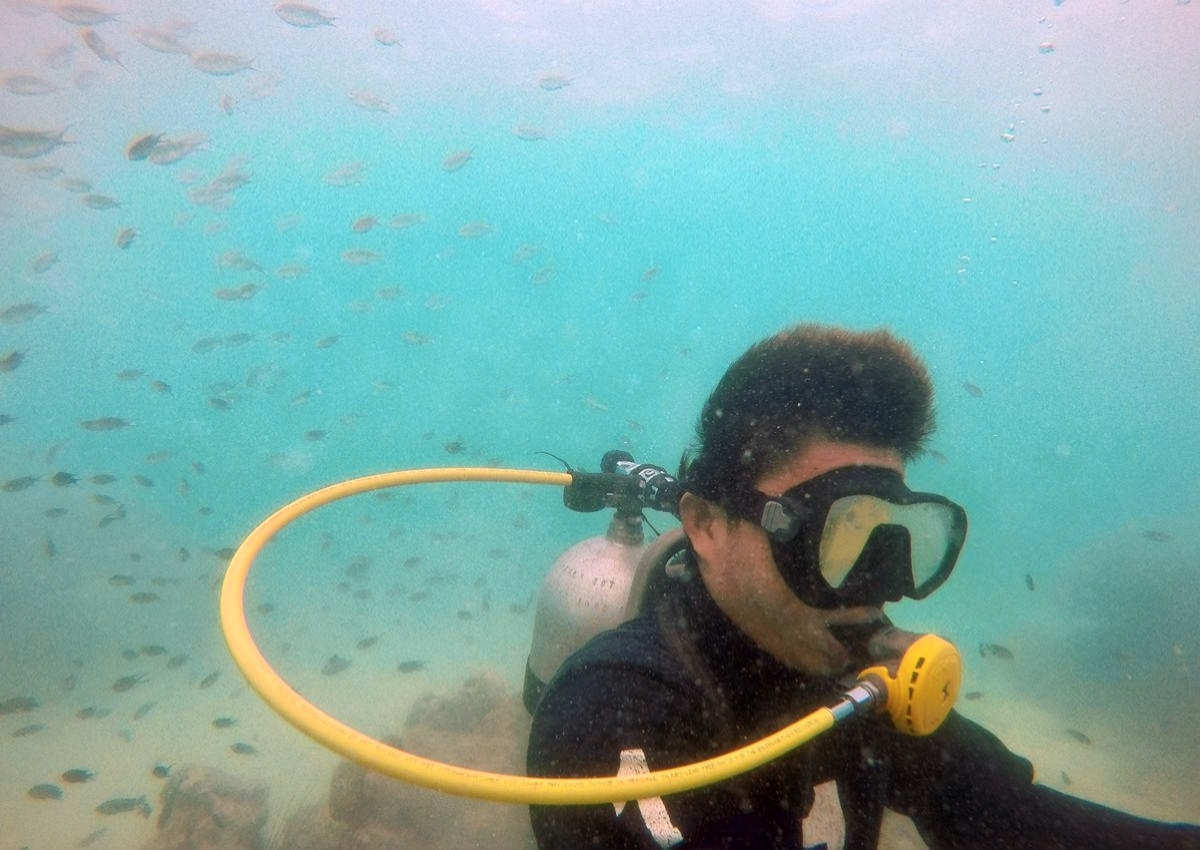}\hspace{0.01\textwidth}
    \includegraphics[width=0.18\textwidth]{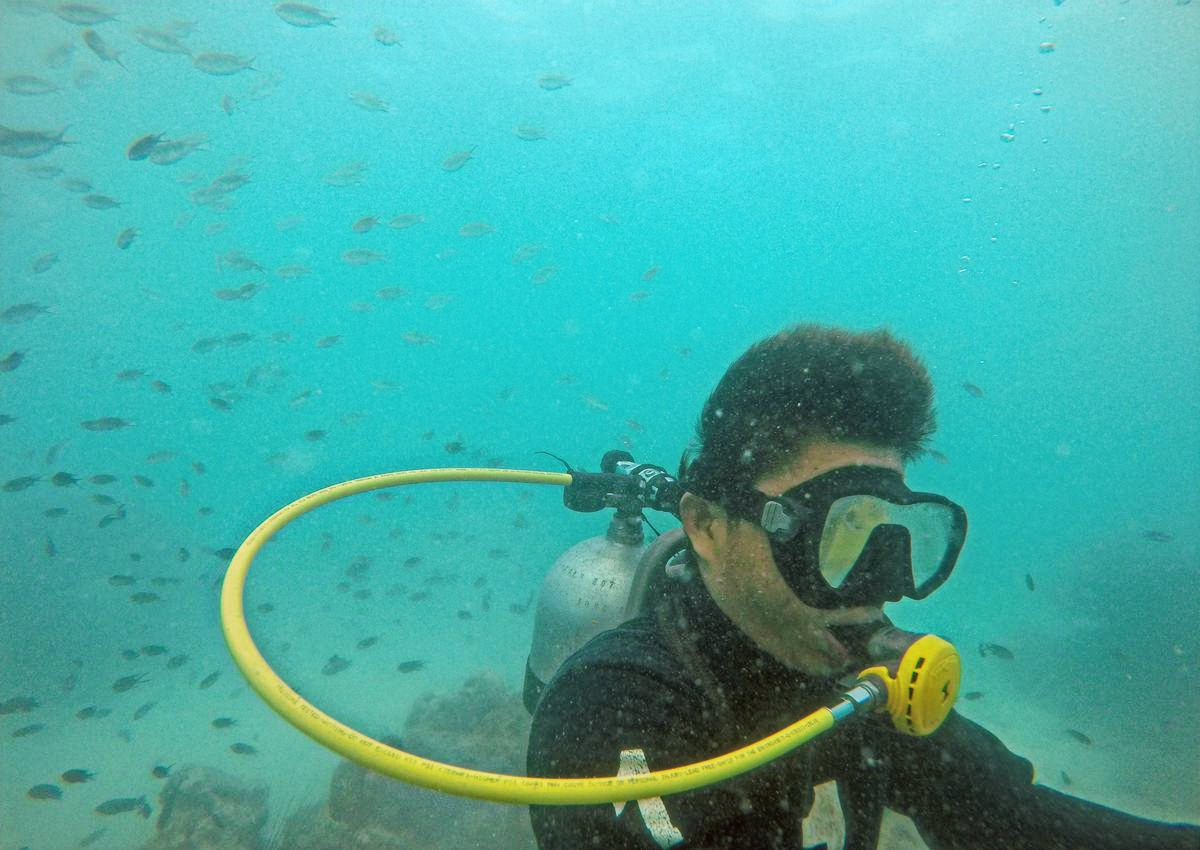}\hspace{0.01\textwidth}
    \includegraphics[width=0.18\textwidth]{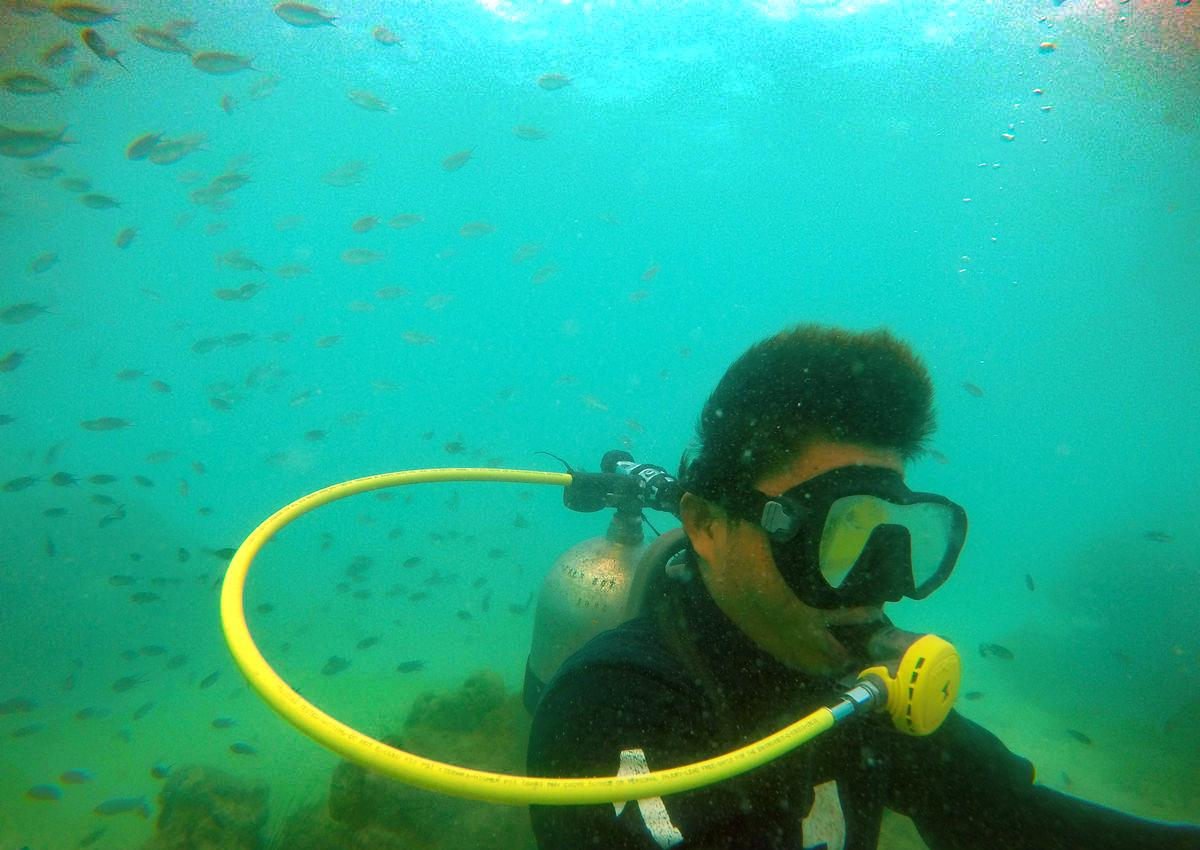}

    \vspace{0.5em}

    \includegraphics[width=0.18\textwidth]{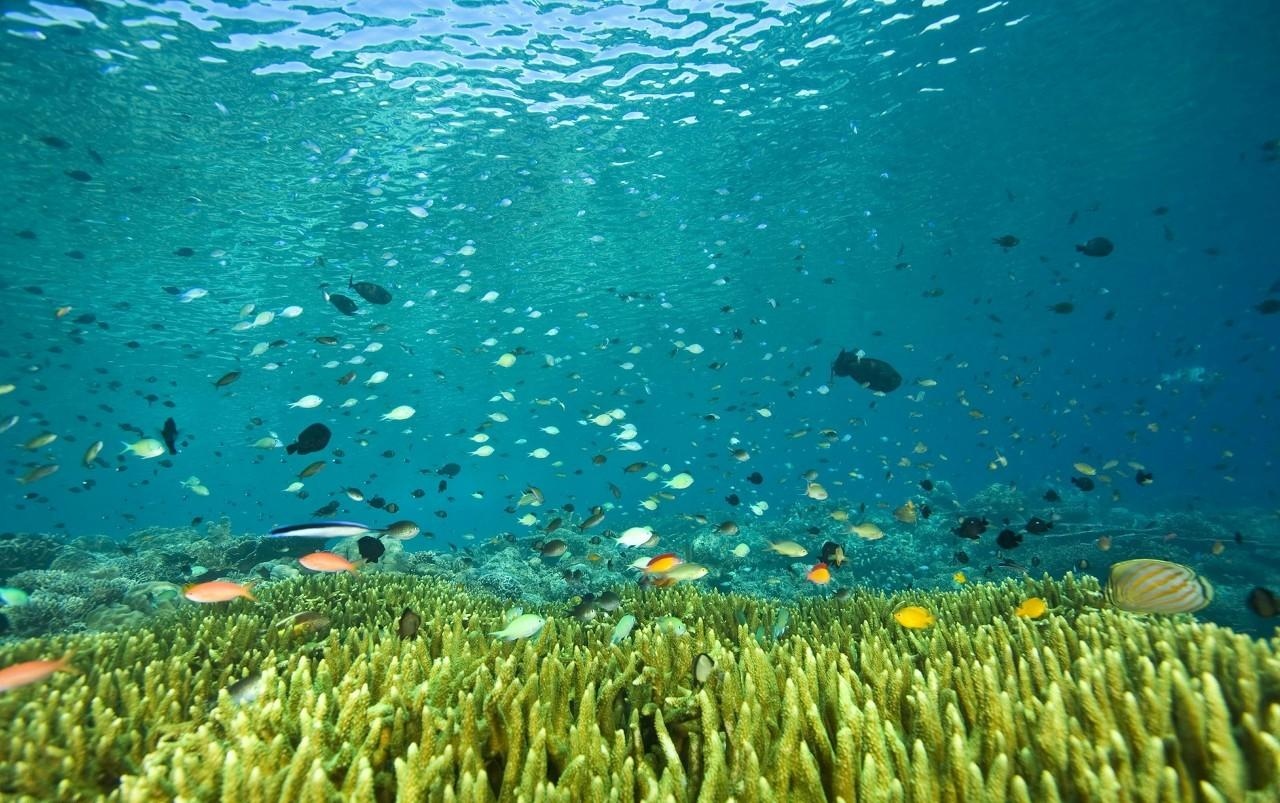}\hspace{0.01\textwidth}
    \includegraphics[width=0.18\textwidth]{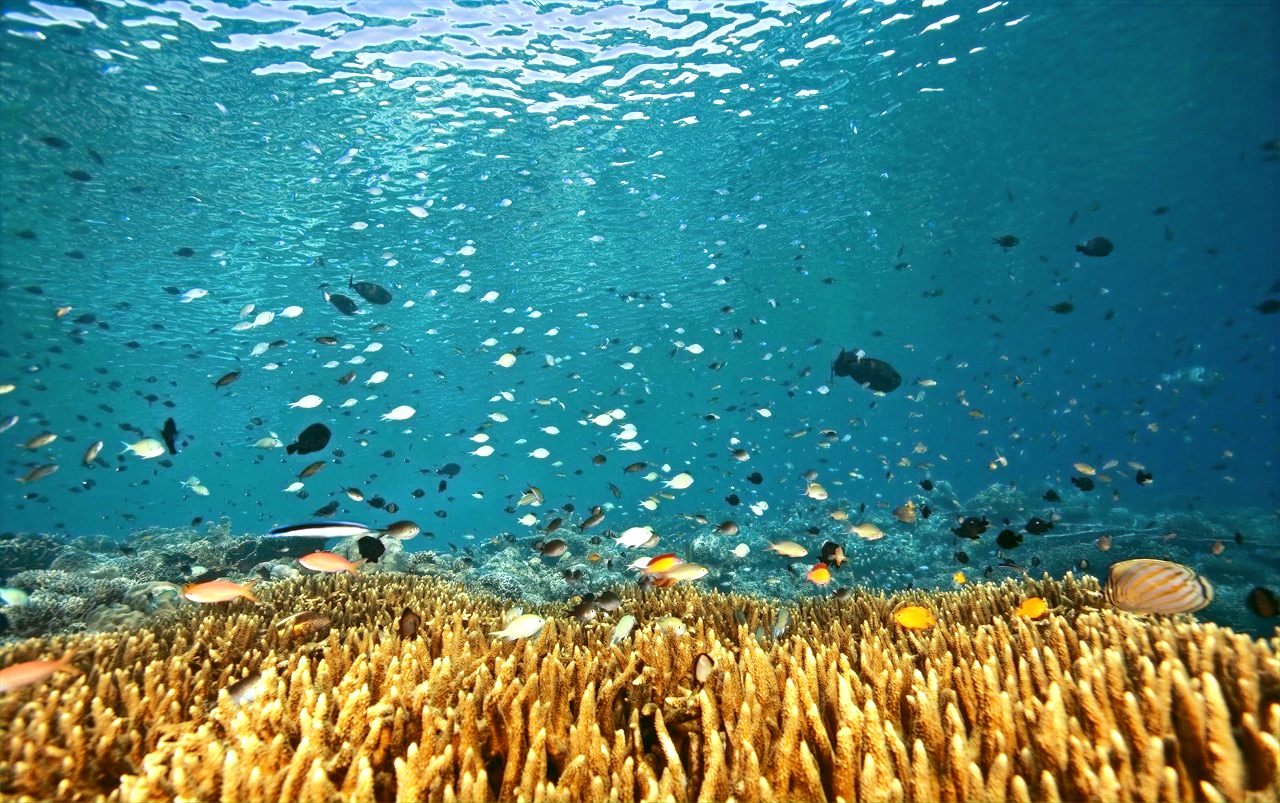}\hspace{0.01\textwidth}
    \includegraphics[width=0.18\textwidth]{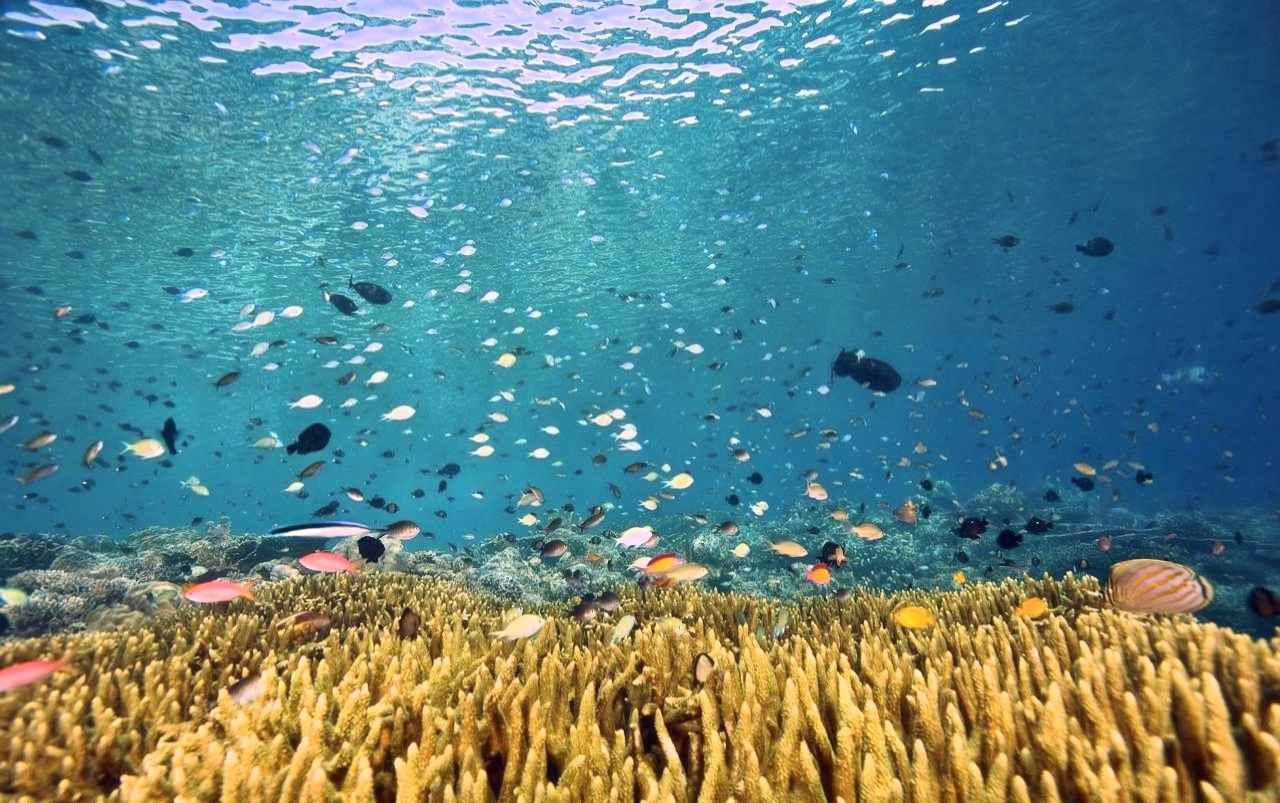}\hspace{0.01\textwidth}
    \includegraphics[width=0.18\textwidth]{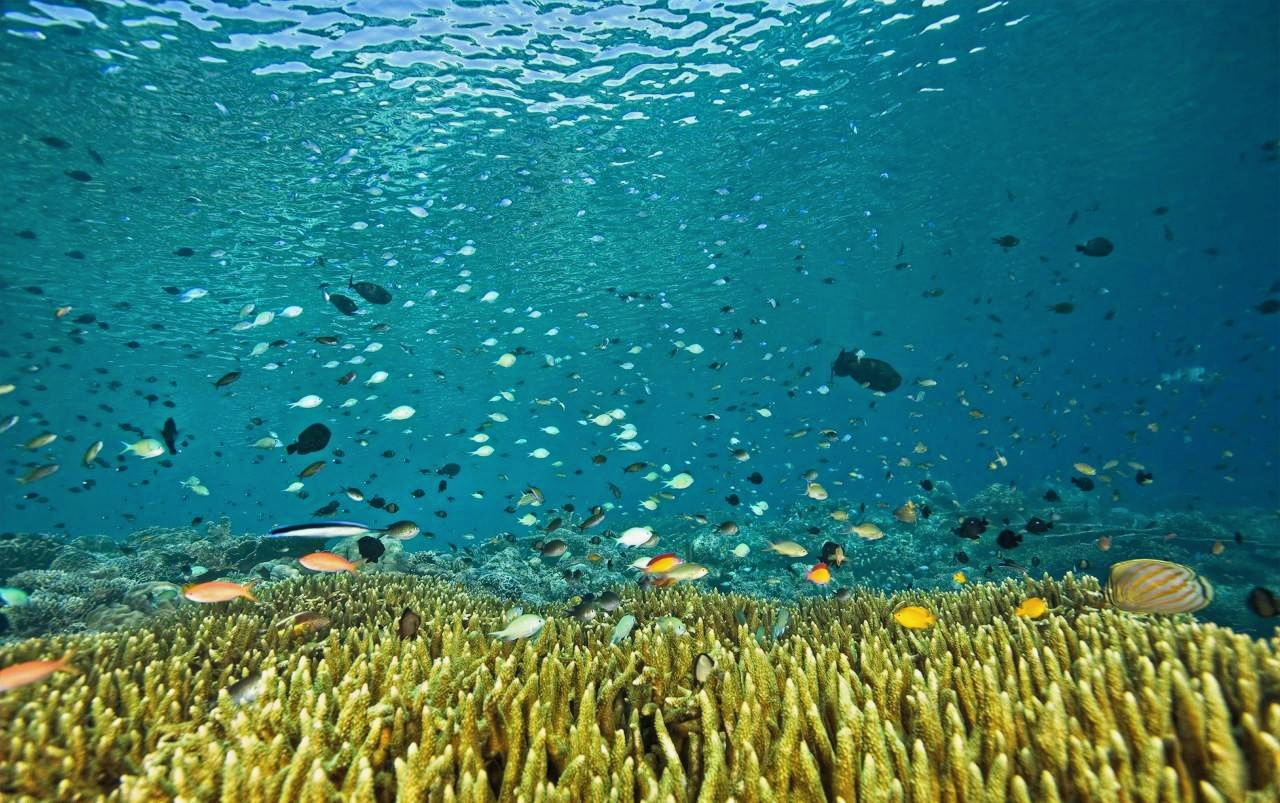}\hspace{0.01\textwidth}
    \includegraphics[width=0.18\textwidth]{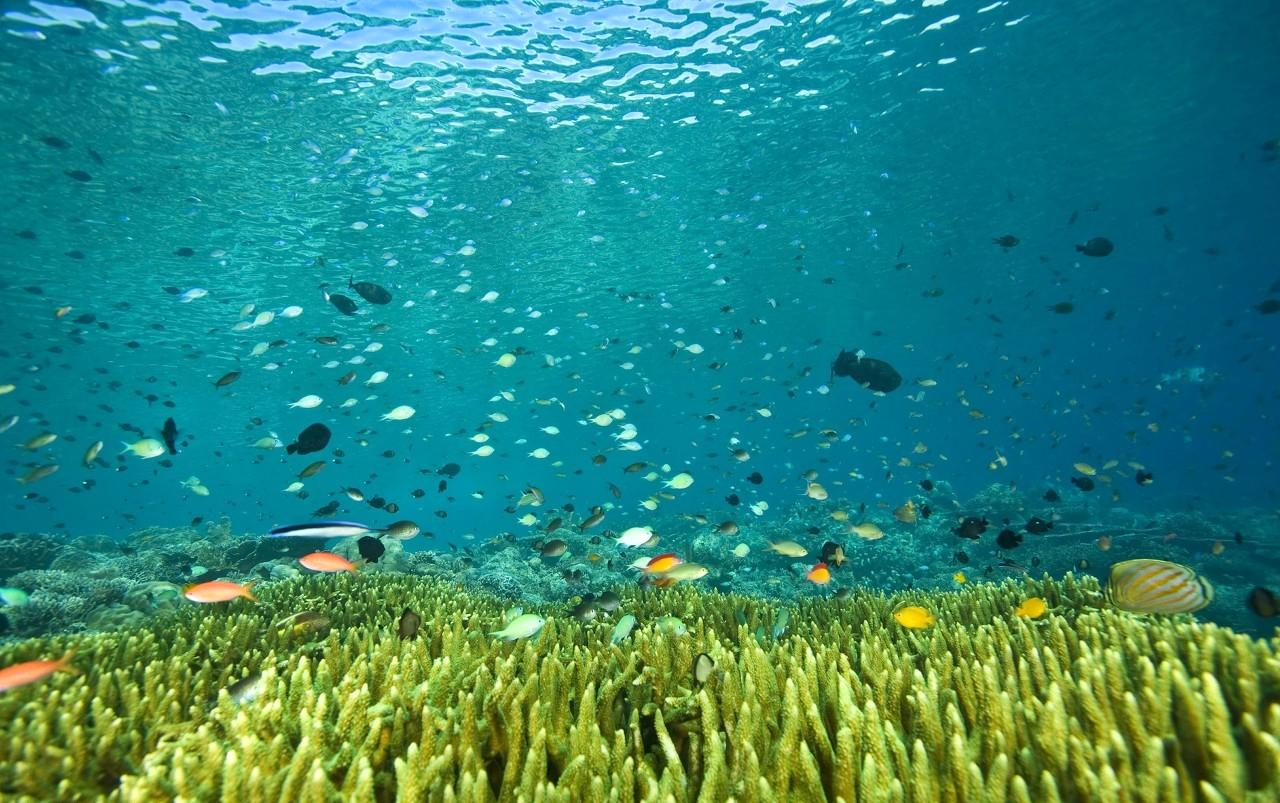}

    \vspace{0.5em}

    \includegraphics[width=0.18\textwidth]{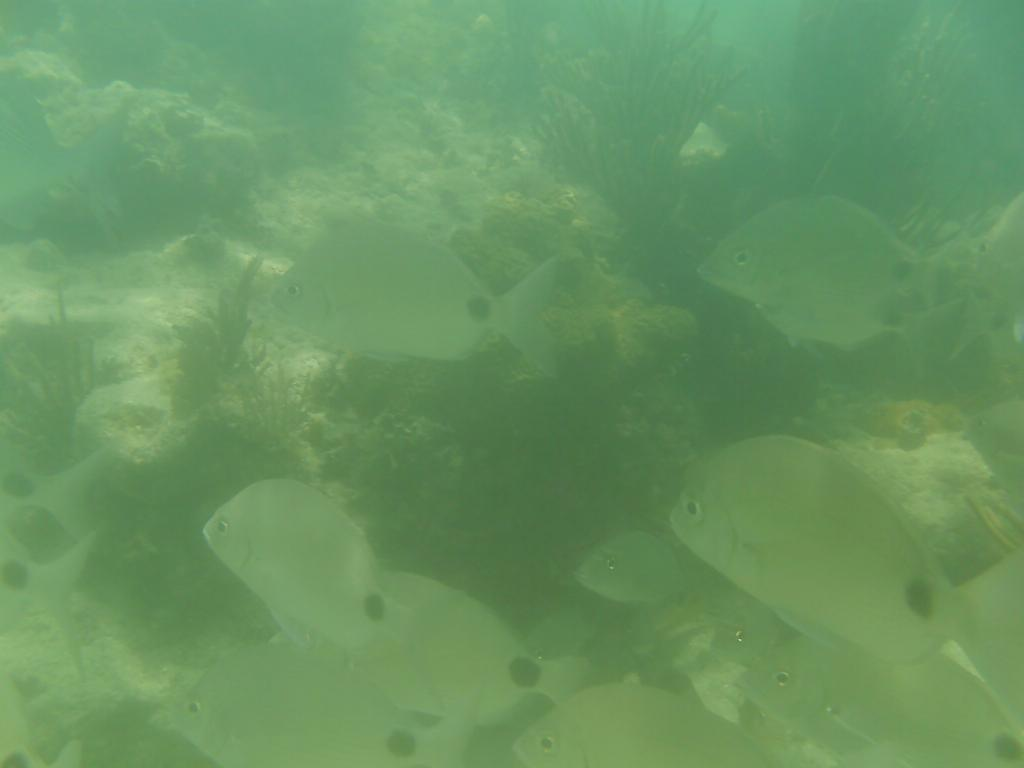}\hspace{0.01\textwidth}
    \includegraphics[width=0.18\textwidth]{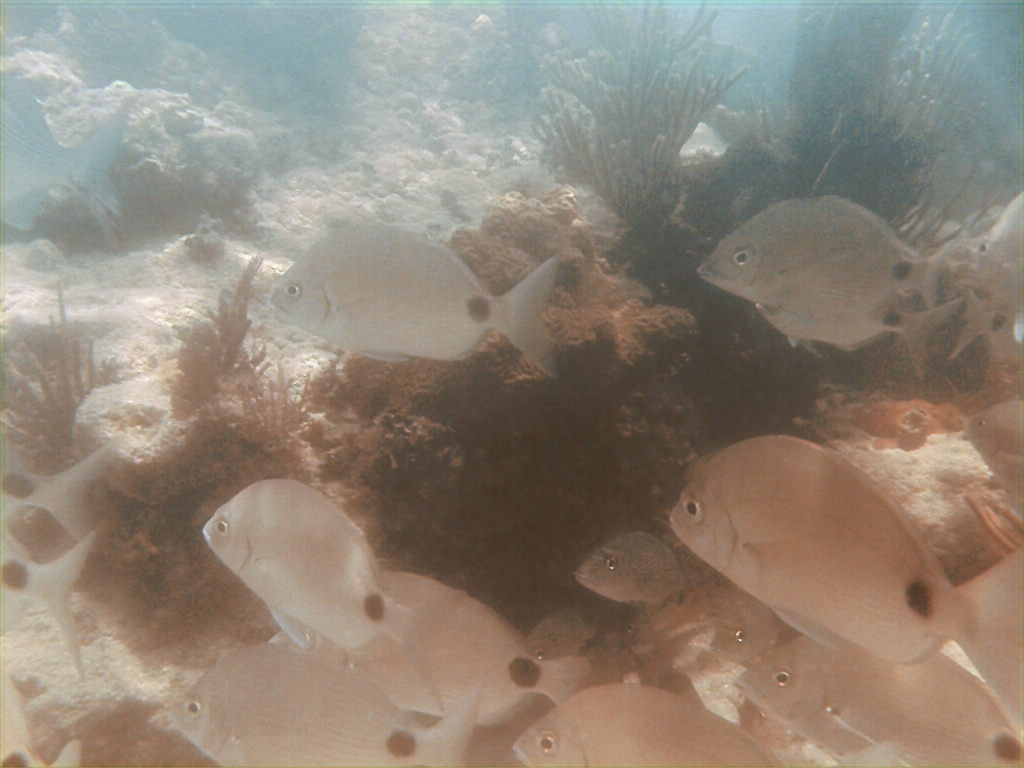}\hspace{0.01\textwidth}
    \includegraphics[width=0.18\textwidth]{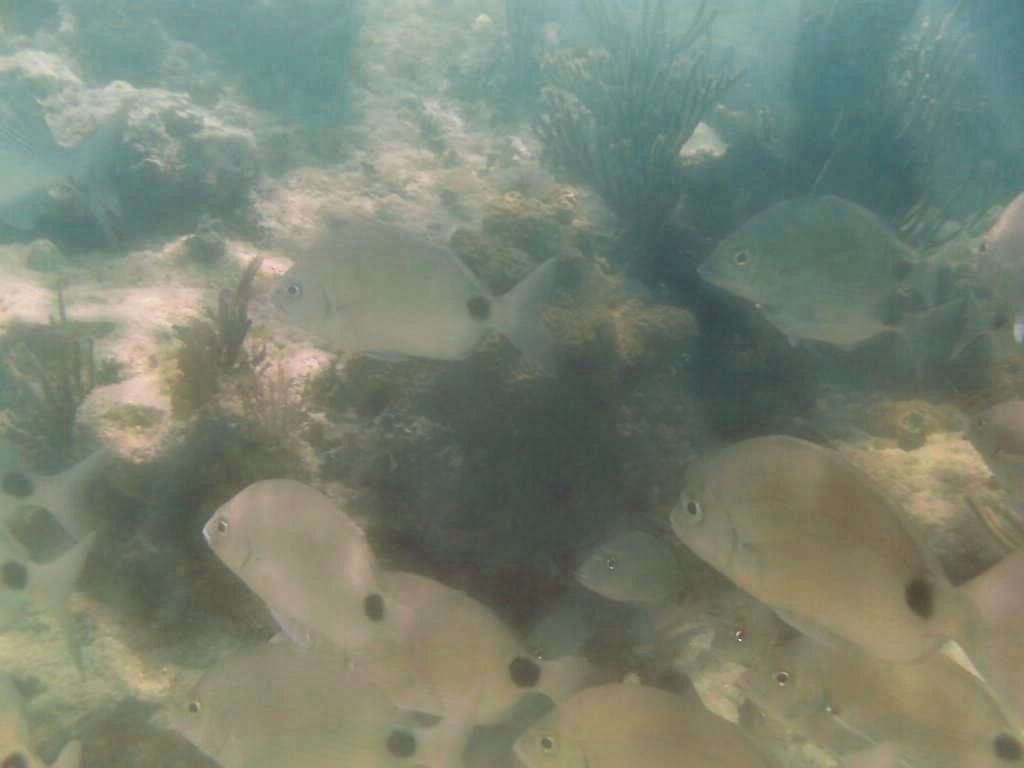}\hspace{0.01\textwidth}
    \includegraphics[width=0.18\textwidth]{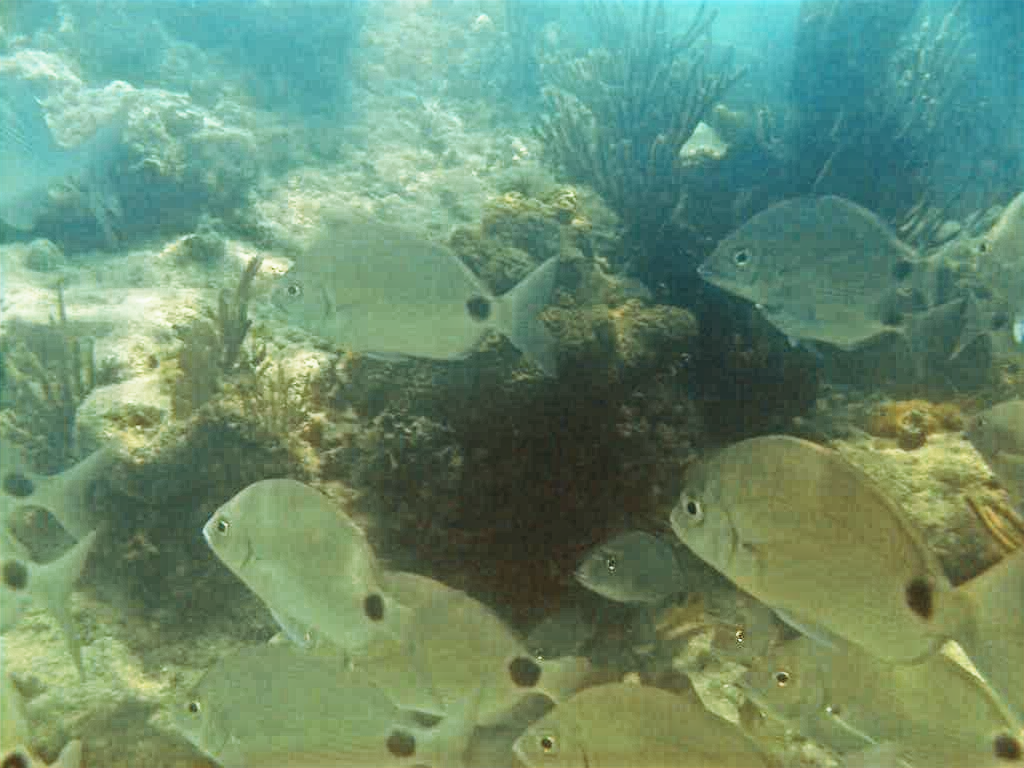}\hspace{0.01\textwidth}
    \includegraphics[width=0.18\textwidth]{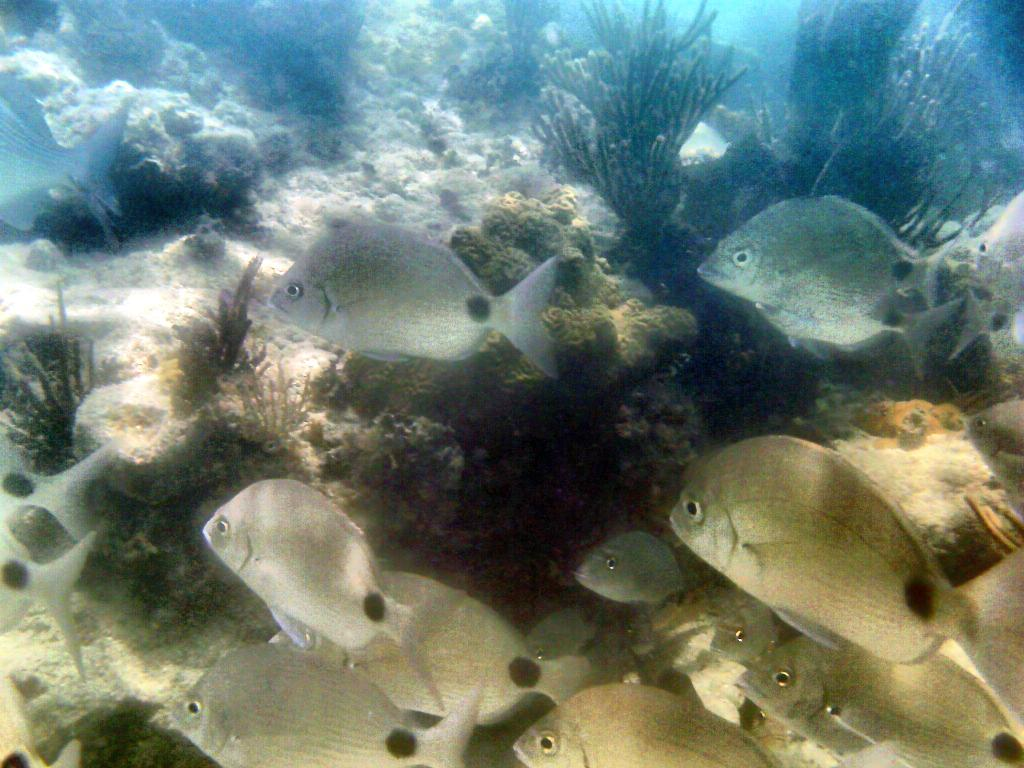}

    \caption{Visual comparison of underwater image enhancement results on representative samples from the UIEB dataset. Columns correspond to the input image, Chen~\emph{et al.}, SCNet, the proposed method, and the ground truth. The proposed method achieves improved color correction, enhanced contrast, and better preservation of structural details across diverse underwater scenes.}
    \label{fig:comparison_visual}
\end{figure*}

\begin{figure*}[!h]
    \centering

    \includegraphics[width=0.18\textwidth]{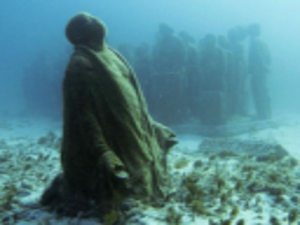}\hspace{0.01\textwidth}
    \includegraphics[width=0.18\textwidth]{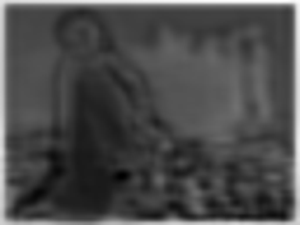}\hspace{0.01\textwidth}
    \includegraphics[width=0.18\textwidth]{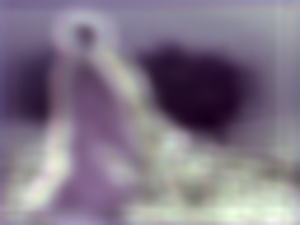}\hspace{0.01\textwidth}
    \includegraphics[width=0.18\textwidth]{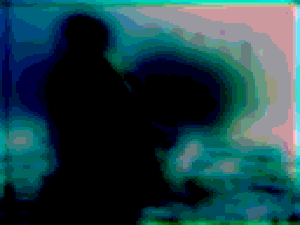}\hspace{0.01\textwidth}
    \includegraphics[width=0.18\textwidth]{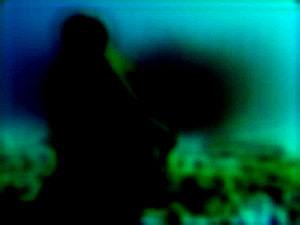}


    \makebox[0.18\textwidth][c]{(a)}
    \hspace{0.007\textwidth}
    \makebox[0.18\textwidth][c]{(b)}
    \hspace{0.007\textwidth}
    \makebox[0.18\textwidth][c]{(c)}
    \hspace{0.007\textwidth}
    \makebox[0.18\textwidth][c]{(d)}
    \hspace{0.007\textwidth}
    \makebox[0.18\textwidth][c]{(e)}

    \vspace{0.4em}

    \includegraphics[width=0.18\textwidth]{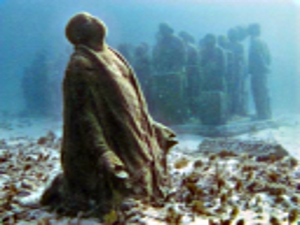}\hspace{0.01\textwidth}
    \includegraphics[width=0.18\textwidth]{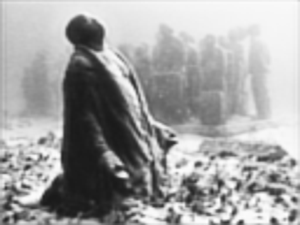}\hspace{0.01\textwidth}
    \includegraphics[width=0.18\textwidth]{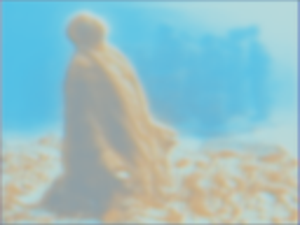}\hspace{0.01\textwidth}
    \includegraphics[width=0.18\textwidth]{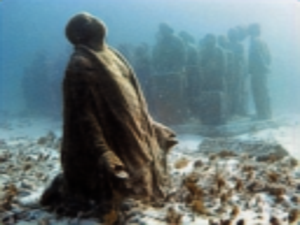}\hspace{0.01\textwidth}
    \includegraphics[width=0.18\textwidth]{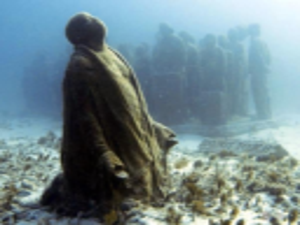}


    \makebox[0.18\textwidth][c]{(f)}
    \hspace{0.007\textwidth}
    \makebox[0.18\textwidth][c]{(g)}
    \hspace{0.007\textwidth}
    \makebox[0.18\textwidth][c]{(h)}
    \hspace{0.007\textwidth}
    \makebox[0.18\textwidth][c]{(i)}
    \hspace{0.007\textwidth}
    \makebox[0.18\textwidth][c]{(j)}

    \caption{Results demonstrating the intermediate outputs for a representative sample from the UIEB dataset. The images denote (a) input, (b) depth map, (c) transmission map, (d) normalized turbidity, (e) normalized noise, (f) estimated scene radiance, (g) illumination, (h) reflectance, (i) final enhanced output image, and (j) ground truth.}
    \label{fig:intermediate}
\end{figure*}

Figure~\ref{fig:comparison_visual} presents a qualitative comparison on three underwater scenes from the UIEB test set, selected to represent varying degradation conditions. Each row corresponds to a different scene.
In the \textit{top row}, SCNet fails to enhance the water-bed sufficiently, while Chen~\emph{et al.} over-enhances it, resulting in an unnatural appearance. In contrast, the proposed method produces a more balanced color appearance while preserving fine structural details. The overall color distribution and contrast achieved by the proposed approach are more consistent with the ground truth, especially in regions with depth variations.
The \textit{second row} shows a wide underwater scene with rich color content. Existing methods tend to alter the color distribution, resulting in noticeable deviations from natural tones. The proposed method, by leveraging physics-guided dehazing and Retinex-based decomposition, maintains more consistent and balanced colors across the scene.
The \textit{bottom row} presents a scene with underwater structures and fish in highly turbid and challenging lighting conditions, where low contrast limits visibility of fine details. While the competing methods achieve some degree of color correction, the overall contrast remains insufficient. The proposed method provides a clearer improvement in contrast and details.

Fig.~\ref{fig:intermediate} illustrates the intermediate outputs of our pipeline on a representative sample from the UIEB dataset. The depth map (b) captures the spatial structure of the scene, while the transmission map (c) reflects the attenuation of light through the water medium. The normalized turbidity map (d) and normalized noise map (e) are visualized after scaling due to their inherently low magnitudes; the turbidity values are near-zero, whereas the noise map retains a discernible pattern. The estimated scene radiance (f) is decomposed into illumination (g), and reflectance (h) through the Stage-2 network. The final output (i) demonstrates a marked improvement over the input, closely approximating the ground truth (j).

\section{Conclusion}
\label{sec:conclusion}

This work proposed a three-stage UIE network that combines physics-based modeling and deep learning. Experiments on the UIEB and UFO datasets demonstrate that the proposed method achieves competitive performance in terms of PSNR, SSIM, and LPIPS. By improving visibility, color fidelity, and structural clarity, this work offers a practical solution for real-world applications such as marine species monitoring, underwater robotics, and environmental analysis.

\bibliographystyle{IEEEbib}
{\fontsize{10}{11.5}\selectfont
\bibliography{ref}}

\end{document}